\documentclass[trackchanges,twocolumn]{aastex701}

\usepackage{wrapfig}
\usepackage{subfloat}
\usepackage{subfig}
\usepackage{appendix}

\begin{document}

\title{The Debris Disk Host $\beta$ Piscis Austrinus is a Rapidly Rotating Star Seen Nearly Pole-On}

\author[orcid=0009-0005-9132-5779,sname='Kane']{Colin Kane}
\affiliation{Department of Physics and Astronomy, Georgia State University, Atlanta, GA 30303, USA}
\email[show]{ckane6@gsu.edu}  

\author[orcid=0000-0001-5313-7498, sname='White']{Russel White} 
\affiliation{Department of Physics and Astronomy, Georgia State University, Atlanta, GA 30303, USA}
\email{rwhite31@gsu.edu}  

\author[0000-0003-3045-5148,sname='Jones']{Jeremy Jones}
\affiliation{Department of Physics and Astronomy, Georgia State University, Atlanta, GA 30303, USA}
\affiliation{The CHARA Array of Georgia State University, Mount Wilson Observatory, Mount Wilson, CA 91023, USA}
\email{jjones176@gsu.edu}  

\author[0000-0002-7982-2095,sname='Montesinos']{Benjamin Montesinos}
\affiliation{Centro de Astrobiolog\'{\i}a (CAB) CSIC-INTA, Villanueva de la Ca\~nada Madrid, Spain}
\email{bmm@gcab.inta-csic.es}  

\author[0009-0006-9244-3707,sname='Carrazco-Gaxiola']{Sebasti\'an Carrazco-Gaxiola}
\affiliation{Department of Physics and Astronomy, Georgia State University, Atlanta, GA 30303, USA} \affiliation{RECONS Institute, Chambersburg, PA 17201, USA}
\email{jcarrazcogaxiola1@gsu.edu}  

\author[0009-0006-4398-4654,sname='Johns']{Tim Johns}
\affiliation{Department of Physics and Astronomy, Georgia State University, Atlanta, GA 30303, USA} \affiliation{RECONS Institute, Chambersburg, PA 17201, USA}
\email{tjohns6@gsu.edu}  

\author[0000-0002-9811-5521,sname='Kar']{Aman Kar}
\affiliation{Department of Physics and Astronomy, Georgia State University, Atlanta, GA 30303, USA} \affiliation{RECONS Institute, Chambersburg, PA 17201, USA}
\email{akar5@gsu.edu}  

\author[0000-0003-0193-2187,sname='Jao']{Wei-Chun Jao}
\affiliation{Department of Physics and Astronomy, Georgia State University, Atlanta, GA 30303, USA} 
\email{wjao@gsu.edu}  

\author[orcid=0000-0002-9061-2865, sname='Henry']{Todd Henry} 
\affiliation{RECONS Institute, Chambersburg, PA 17201, USA}
\email{thenry88@gsu.edu}

\begin{abstract}

Previous studies of $\beta$ Piscis Austrinus (PsA) have speculated that the narrow and saddle-like shapes of some of its weak metallic lines are a consequence of it being a rapidly rotating star viewed nearly pole-on. Here we use the \texttt{fastrot-spec} spectral synthesis code to model high-dispersion (R = 115,000) HARPS spectra of $\beta$ PsA in order to determine its inclination and photospheric properties, with additional constraints on the surface temperature set by measures of \ion{Fe}{2}/\ion{Fe}{1} line ratios. The analysis confirms that $\beta$ PsA is oriented nearly pole-on ($i = 4.75^{+0.75}_{-0.50}$~$^o$) and experiences substantial gravity darkening caused by its rapid rotation ($\Omega/\Omega_{crit}=0.93\pm0.17$). $\beta$ PsA has a polar temperature of $10300^{+200}_{-250}$~K that is 24\% hotter than its equatorial temperature ($8275^{+317}_{-400}$~K). This results in its apparent luminosity being 48\% larger than its actual luminosity of 26.2$^{+1.9}_{-2.4}$~L$_\odot$. When this methodology is applied to high-dispersion spectra of the star Vega, the analysis determines a nearly pole-on orientation that is consistent with interferometric measurements, validating the technique. Based on comparisons with PARSEC evolutionary models of stars rotating at similar velocities, 
$\beta$ PsA has a mass of $2.20\pm0.03$~M$_{\odot}$ and an age of $141^{+113}_{-49}$~Myr; this age is consistent with the age inferred for its G5V companion star, CD-32 17127, based on lithium depletion models.  The analysis demonstrates the potential for both identifying and determining the stellar properties of rapidly rotating stars viewed nearly pole-on via spectroscopy alone.

\end{abstract}

\section{Introduction}\label{sec:intro} 

The star $\beta$ Pisces Austrinus ($\beta$~PsA = 17~PsA = HD~213398 = HR~8576) is a nearby (D $\simeq45$~pc; \citealt{gaia_collaboration_gaia_2023}), bright (V = 4.29), southern hemisphere (DEC = $-32^{\circ}$), spectral type A1Va star \citep{gray_early_1987} with solar composition ([Fe/H] = 0.01; \citealt{soubiran_pastel_2016}).  It exhibits sharp absorption lines because of its relatively low projected rotational velocity (\textit{v}~sin~\textit{i}~=~26~$\mathrm{km~s}^{-1}$; \citealt{glebocki_systematic_2005}). Its well-defined spectral lines have made it amenable to detailed spectroscopic studies, including abundance anomaly analyses \citep{holweger_high-resolution_1986, lemke_abundance_1989, lemke_abundance_1990}, physical parameter studies \citep{fitzpatrick_determining_1999}, and high-precision radial velocity searches for low mass companions \citep{lagrange_extrasolar_2009}. 

Based on high dispersion optical spectra (R $\simeq$ 62,500), \cite{gulliver_metallic_1991} noted that some of the weak metallic lines of $\beta$ PsA have peculiar profiles similar to those seen in the low \textit{v}~sin~\textit{i} star Vega ($\alpha$ Lyrae), with spectral type A0V \citep{gray_contributions_2003}. \cite{takeda_determination_2021} assembles seven previous \textit{v}~sin~\textit{i} measurements for Vega (see their Table 1), with a mean of 21.4~$\mathrm{km~s}^{-1}$. The peculiar profiles are caused by temperature differences across the photosphere, attributed to gravity darkening from rapid rotation and its nearly pole-on orientation. Gravity darkening, the cooling of the equatorial regions of a star distorted by rapid rotation \citep{von_zeipel_radiative_1924}, alters the ionization fractions from the pole to the equator, causing the integrated line profile to become inclination-dependent \citep{montesinos_surface_2024}. This phenomenon appears to be most evident in weak metallic lines that are temperature sensitive. This has been extensively studied in the prototype for these distorted profiles, Vega \citep{gulliver_vega_1994, hill_spectral_2004, yoon_effect_2008, takeda_rotational_2008, hill_study_2010, yoon_new_2010, takeda_determination_2021, montesinos_surface_2024}. 

Using modern prescriptions for gravity darkening profiles, \cite{montesinos_surface_2024} successfully modeled the peculiar line profiles of Vega. Figure \ref{fig:dist_v_nondist} below shows one of the distorted line profiles in the spectrum of $\beta$ PsA, which we describe as ``saddle-shaped" due to the upturn in the line center (e.g., \ion{Ba}{2} $\lambda 4554~ \rm{\AA}$), as well as a ``normal" or ``bowl-shaped" (e.g., \citealt{gray_observation_2008}) absorption line (e.g., \ion{Cr}{2} $\lambda 4555~ \rm{\AA}$), highlighting how only temperature-sensitive lines show this line distortion. We note that $\beta$ PsA and Vega have similar spectral types, effective temperatures, and \textit{v}~sin~\textit{i} values.

Far Infrared and submillimeter observations from the \textit{Herschel} DEBRIS survey \citep{phillips_target_2010,thureau_unbiased_2014} identify $\beta$ PsA as a debris disk host based on its far infrared excess. \cite{pearce_planet_2022} model this disk, finding it to extend out to $60 \pm 30$ AU. Because the disk is unresolved in the \textit{Herschel} observations and not yet imaged with other facilities, its inclination is unknown.

$\beta$ PsA has a lower mass stellar companion (CD-32 17127) separated by 30\farcs33 and is spectroscopically classified as a G5V \citep{torres_search_2006}. Both stars share common proper motion \citep{waisberg_binarity_2023} and have a projected separation of 1373 AU; the orbit is undetermined.  \citet{vican_age_2012} estimate an age of 180 Myr for $\beta$ PsA based on comparisons with YREC stellar evolutionary models \citep{pinsonneault_distances_2004}. This youthful age is consistent with the detection of lithium in the spectrum of CD-32 17127 \citep{torres_search_2006}. It is not associated with any nearby moving groups, based on kinematic membership probabilities in \cite{gagne_montreal_2026}.

In this paper, we model the spectrum of $\beta$ PsA in order to determine its inclination and stellar properties. Section \ref{sec:spectra} discusses the spectroscopic data, including the reduction and continuum normalization steps. Section \ref{sec:modeling} discusses how the spectral models are generated and compared to the HARPS spectrum of $\beta$ PsA, while also describing how we determine our best-fit model parameters. Section \ref{sec:chiron} validates our results using a CHIRON spectrum of $\beta$ PsA. Section \ref{sec:LandT} discusses the impact gravity darkening has on luminosity and temperature calculations. Section \ref{sec:evolution} describes how the luminosity, mass, and age of the star are determined. Section \ref{sec:vega} demonstrates the robustness of this methodology by applying it to Vega. Section \ref{sec:discussion} presents a summary and a discussion of the implications of this project on future work.

\section{Spectroscopic Observation and Reduction}\label{sec:spectra} 

\subsection{HARPS Spectra of $\beta$ PsA}\label{subsec:HARPS}

High-dispersion, high-resolution spectra from the HARPS (High-Accuracy Radial Velocity Planetary Searcher; \citealt{mayor_setting_2003}) echelle spectrograph are used in this study. HARPS is mounted on the 3.6-m telescope at the La Silla Observatory and it covers a spectral range of 3782~\AA\ to 6913~\AA\, with a resolving power of 115,000. A total of 22 HARPS optical spectra of $\beta$ PsA are obtained from the European Southern Observatory (ESO) archive\footnote{https://archive.eso.org/scienceportal/home} \citep{european_southern_observatory_eso_harps_2014}. These spectra were reduced, calibrated, and merged via the ESO pipelines. The pipeline does barycentric corrections and stitches the echelle orders together to form one continuous spectrum. The spectra have signal-to-noise (S/N) values between 129 and 280, with a mean of 210. The spectra of $\beta$ PsA were stacked by summation of counts at each discrete wavelength, resulting in one very high S/N spectrum (S/N $\sim$ 980), ideal for studying weak line profiles.


Continuum normalization was conducted using the \texttt{mdwarf\_contin} Python code \citep{medan_importance_2025}. \texttt{mdwarf\_contin} calculates the \textit{alpha shape} \citep{edelsbrunner_shape_1983} of the data and uses a local polynomial regression fitting algorithm to determine the continuum (for a more in-depth description, see Section 3.1 of \citealt{medan_importance_2025}). This code was designed to identify the pseudo-continua of M dwarf spectra, but the original technique was made to empirically model the echelle blaze and stellar continuum \citep{xu_modeling_2019,cretignier_rassine_2020}.

Finally, the spectrum of $\beta$ PsA was Doppler-shifted to correct for the star's radial velocity. The Doppler shift is calculated by cross-correlating the observed spectrum and a synthetic spectrum used in this analysis (see Section \ref{sec:modeling}), where the peak of the cross-correlation function was adopted as the Doppler shift using the \texttt{PyAstronomy}\footnote{https://github.com/sczesla/PyAstronomy} package \citep{czesla_pya_2019}. Using this method, we determine a radial velocity of $\beta$ PsA to be 5.26$\pm0.10$~km~s$^{-1}$, which is consistent with previous results (e.g., 5.14$\pm0.50$~km~s$^{-1}$; \citealt{gaia_collaboration_gaia_2023}). The uncertainty is set to a conservative estimate of the absolute radial-velocity zero point of HARPS, accounting for wavelength calibration systematics at the tens of m~s$^{-1}$ level \citep{lo_curto_harps_2015,cersullo_new_2019}.

\subsection{CHIRON Spectra of $\beta$ PsA and CD-32 17127}\label{subsec:CHIRON}

A spectrum of $\beta$ PsA was obtained using the SMARTS/CTIO 1.5-m telescope in Cerro Tololo, Chile, with the CHIRON echelle spectrograph \citep{tokovinin_chironfiber_2013} on 26 August 2025. CHIRON covers a wavelength range of 4500~\AA\ to 8900~\AA\, spanning 62 orders; the spectrum was obtained in slit mode, which has a resolving power of $\sim95,000$. The spectrum has an average S/N of 155 across the orders used in this analysis ($\lambda4504.6$~\AA\ to $\lambda4952.6$~\AA). A CHIRON spectrum of the companion to $\beta$ PsA, CD-32~17127, was obtained on 27 December 2025 using slicer mode, which yields a resolving power of $\sim80,000$. This observing mode was used for CD-32 17127 because there are a wide variety of comparison stars observed with this setup (e.g., \citealt{paredes_solar_2021}; \citealt{nisak_mapping_2022}; \citealt{hubbard-james_solar_2026}). 

Thorium-Argon lamp spectra were obtained immediately following the observations for wavelength calibration. The data were reduced using the standard spectral reduction pipeline \citep{tokovinin_chironfiber_2013,paredes_solar_2021}. The same prescription for normalizing the continuum used for the HARPS spectrum is applied to the CHIRON spectra (see Section \ref{subsec:HARPS}). The spectra were barycentric corrected using  \texttt{radial\_velocity\_correction} function from the \texttt{Astropy} python package \citep{astropy_collaboration_astropy_2013, astropy_collaboration_astropy_2018, astropy_collaboration_astropy_2022} and are Doppler shifted using the same radial velocity as the HARPS spectrum (see Section \ref{subsec:HARPS}).

\subsection{HIDES spectra of Vega}\label{subsec:HIDES}

Finally, for a comparison analysis, we used the high resolution (R $\sim100,000$), high S/N ($\gtrsim2000$) optical spectrum of Vega provided by \cite{takeda_high-resolution_2007}. The spectrum was obtained with the HIgh-Dispersion Echelle Spectrograph (HIDES; \citealt{izumiura_hides_1999}) on the 188-cm Okayama Astrophysical Observatory telescope. HIDES covers the wavelength ranges of 3900 \AA\ to 7200 \AA\ and 7600 \AA\ to 8800 \AA. The spectrum from \cite{takeda_high-resolution_2007} has been velocity corrected and normalized. To remain consistent, we renormalized the spectrum using the same method applied to the HARPS and CHIRON spectra.

\section{Modeling the Spectrum of $\beta$ PsA}\label{sec:modeling} 

\subsection{The Spectral Synthesis Code \rm{\texttt{fastrot-spec}}}

A set of metallic line profiles is modeled using the spectral synthesis code \texttt{fastrot-spec}\footnote{The code and the affiliated files required are available at \url{https://github.com/astrobmm/fastrot-spec}} \citep{montesinos_fastrot-spec_2024}. This code is specifically designed to build the spectrum of rapidly rotating stars with a radiative envelope, viewed at any inclination. \texttt{fastrot-spec} first carries out a parameterization of the stellar surface, providing the values of the effective temperature ($T_{\rm eff}$), surface gravity ($\log g_{\rm eff}$), and radius ($r$), as functions of the latitude. This parameterization is based on the work by \cite{espinosa_lara_gravity_2011}. The photosphere is then divided into cells delimited by intersections of a fine mesh of meridians and parallels. Once this has been set up, the emergent spectrum from each cell is computed using the corresponding values of $T_{\rm eff}$ and $\log g_{\rm eff}$.  Taking into account the inclination, $i$, and the equatorial velocity, $v_{\rm eq}$, each individual spectrum is then blue- or red-shifted, with its intensity corrected for limb darkening and weighted by the projected surface area. The final spectrum is obtained by summing the contributions of all visible cells. The individual spectra are computed using the codes \texttt{ATLAS} and \texttt{SYNTHE} \citep{niemczura_model_2014}, and the model atmospheres by \cite{castelli_new_2003}.

\begin{figure}
    \centering
    \includegraphics[width=\linewidth]{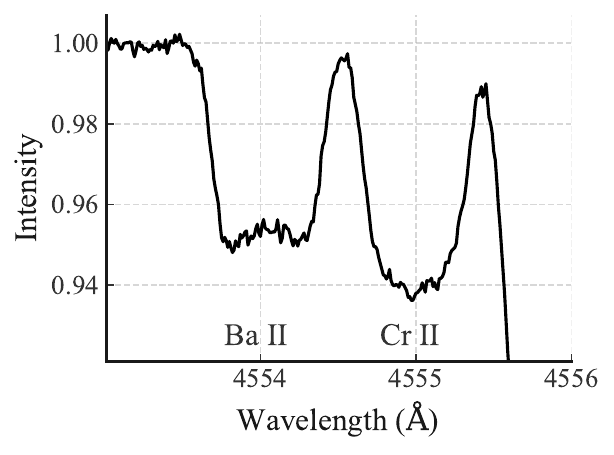}
    \caption{When viewed nearly pole-on, the rapid rotation of $\beta$~PsA causes temperature-sensitive, weak metallic lines to develop saddle-shaped absorption profiles (e.g., \ion{Ba}{2}~$\lambda4554$~\AA). In contrast, less temperature-sensitive lines (e.g., \ion{Cr}{2}~$\lambda4555$~\AA) display bowl-shaped profiles consistent with standard rotational broadening.}
    \label{fig:dist_v_nondist}
\end{figure}

Not all stellar absorption lines are effective in this analysis, as only temperature-sensitive metallic lines show significant distortion of their profiles (see Figure~\ref{fig:dist_v_nondist}). \cite{montesinos_surface_2024} identify and use a set of 30 metallic lines to model the spectrum of the star Vega. Of these, we identify a subset of 11 in the HARPS spectrum that show distorted, saddle-like profiles in the interval 4000~\AA\ to 5000~\AA. These lines are \ion{Ca}{1} 4435 \AA, \ion{Ca}{1} 4454 \AA, \ion{Fe}{1} 4068 \AA, \ion{Fe}{1} 4132 \AA, \ion{Fe}{1} 4494 \AA, \ion{Fe}{1} 4528 \AA, \ion{Fe}{1} 4736 \AA, \ion{Ba}{2} 4554 \AA, \ion{Ba}{2} 4934 \AA, \ion{Ni}{1} 4714 \AA, and \ion{Cr}{2} 4891 \AA. 

\subsection{Initial Parameters}\label{subsec:init_params}
 
The \texttt{fastrot-spec} code requires six parameters to compute the synthetic spectrum. These are the inclination ($i$), polar temperature ($T_{\rm{pole}}$), stellar mass ($M_{\rm{star}}$), equatorial radius ($R_{\rm{eq}}$), the fraction of the angular velocity relative to the Keplerian velocity ($\omega$), and the metallicity ([M/H]). The first two of these parameters, the inclination and polar temperature, are the most difficult to determine precisely. We constrain these by comparing the observed spectrum to a grid of inclination and polar temperature values, using adopted values for the other four parameters, some of which are adjusted interactively (i.e., stellar mass). The range of inclinations we test are between 3.75$~^{\mathrm{o}}$ and 7.00$~^{\mathrm{o}}$, in steps of 0.25$~^{\mathrm{o}}$. Models with inclination $< 3.75~^{\circ}$ are not generated; below this limit, sin~\textit{i} becomes sufficiently small that the equatorial velocity, $v_{\mathrm{eq}}$, exceeds the Keplerian velocity of the star, which is defined as the angular velocity where the acceleration due to gravity and equatorial centripetal acceleration are balanced. Inclinations above $\sim7^{~\mathrm{o}}$ do not produce sufficiently distorted lines. The polar temperature range we test is anchored by the effective temperature of the star taken from \cite{zorec_rotational_2012}. The effective temperature should be an approximate lower limit for the polar temperature of the star, since gravity darkening decreases the temperature at the equator of the star, making the average temperature of the star ($\sim {T}_{\mathrm{eff}})$ cooler than the polar temperature. The range of polar temperatures spans 9450 K to 10700 K, in steps of 50 K. 

The stellar mass is set by comparing $\beta$ PsA's luminosity and ``mean-radius effective temperature", as determined by the best-fit stellar spectrum, to the predictions of stellar evolutionary models (see Section \ref{sec:evolution}). The mean-radius effective temperature is defined by Stefan-Boltzmann's Law using the luminosity calculated by \texttt{fastrot-spec} and the mean radius of the oblate star model. We adopt a mass of 2.34 M$_{\odot}$ from \cite{zorec_rotational_2012} to generate initial stellar spectra and stellar parameters, and then adjust the stellar mass iteratively to be consistent with evolutionary model predictions. The analysis converges quickly to a final stellar mass of $2.20 \pm 0.03$ M$_{\odot}$.

The equatorial radius ($R_{\mathrm{eq}}$) is calculated from the angular diameter estimate from the JMMC\footnote{https://www.jmmc.fr/} (Jean-Marie Mariotti Center) Stellar Diameters Catalog (JSDC; $\theta=0.4525\pm0.0428$; \citealt{bourges_jmmc_2014}), and the astrometric parallax from \textit{Gaia} DR3 ($\pi=22.0836\pm0.2117$ mas; \citealt{gaia_collaboration_gaia_2023}). JSDC estimates angular diameters of stars by creating a relation between interferometrically measured angular diameters and their photometric colors, then applies that relation to stars whose diameters have not been measured. The $R_{\mathrm{eq}}$ is computed using the following relation \citep{aufdenberg_first_2006}:

\begin{equation} \label{eq:req2}
    \frac{R_{\mathrm{eq}}}{R_\odot} = 107.48 \frac{\theta_{\mathrm{mas}}}{\pi_{\mathrm{mas}}},
\end{equation}

\noindent which yields $R_{\mathrm{eq}} = 2.20\pm0.21~\mathrm{R}_{\odot}$ for $\beta$ PsA. Once the inclination, mass, and radius are determined, the angular velocity and Keplerian velocity are calculated from the star's projected rotational velocity (\textit{v}~sin~\textit{i}). Using a \textit{v}~sin~\textit{i} value of $26~\mathrm{km}~\mathrm{s}^{-1}$ \citep{glebocki_systematic_2005}, the inclination, and $R_{\mathrm{eq}}$ we determine the angular velocity, $\Omega$, of the star using the relation:

\begin{equation}\label{eq:Omega}
    \Omega = \frac{v~\mathrm{sin}~i/\mathrm{sin}~i}{R_{\mathrm{eq}}}.
\end{equation}

\noindent The Keplerian velocity of the star is computed using Eq. \ref{eq:Omega_k}, taken from \cite{espinosa_lara_gravity_2011},

\begin{equation}\label{eq:Omega_k}
    \Omega_{k} = \sqrt{\frac{GM}{R_{\mathrm{eq}}^3}}.
\end{equation}

\noindent Finally, taking the ratio of the angular velocity and the Keplerian velocity, we find $\omega$:

\begin{equation}\label{eq:omega}
    \omega = \frac{\Omega}{\Omega_{k}};
\end{equation}

\noindent This parameter varies for each model that we compute, since it relies on the equatorial velocity, which changes as the inclination changes when holding the observed $v$~sin~$i$ fixed. 
 
The final input parameter is the metallicity, which is adopted to be solar ([M/H] = 0.0 dex), consistent with the value of [Fe/H] = 0.01 from \cite{soubiran_pastel_2016}. The initial model grid parameters are listed in Table \ref{tab:model_grid}.

\begin{deluxetable}{ccccccc}
\tablecaption{Initial Model Grid Parameters \label{tab:model_grid}}
\tablehead{
  \colhead{Star} & \colhead{$i$}  & \colhead{$T_{\mathrm{pole}}$} & \colhead{$M_{\mathrm{star}}$} & \colhead{$R_{\mathrm{eq}}$} & \colhead{$\omega$} & \colhead{Metallicity} \\
  \colhead{} & \colhead{$^{\circ}$} & \colhead{K} & \colhead{$\mathrm{M}_{\odot}$} & \colhead{$\mathrm{R}_{\odot}$} & \colhead{} & \colhead{[M/H]} 
}
\startdata
  $\beta$ PsA & 3.75 - 7.0 & 9450 - 10700 & 2.34 & 2.20 & 0.883 - 0.474 & +0.0  \\
  Vega & 4.0 - 8.0 & 9600 - 10650 & 2.15 & 2.73 & 0.791 - 0.396 & $-$0.5 \\
\enddata
\tablecomments{The grid steps for the inclination are 0.25 degrees and for the polar temperature the steps are 50 K.}
\end{deluxetable}


\subsection{Optimizing the Spectral Fit}

To model the 11 metallic line profiles, a $46~\mathrm{km}~\mathrm{s}^{-1}$ ($\frac{d\lambda}{\lambda}$ of $3.04\times10^{-4}$) portion of the spectrum, centered on each line, is used; this width ensures that the surrounding continua is included. 

\begin{figure}
    \centering
    \includegraphics[width=\linewidth]{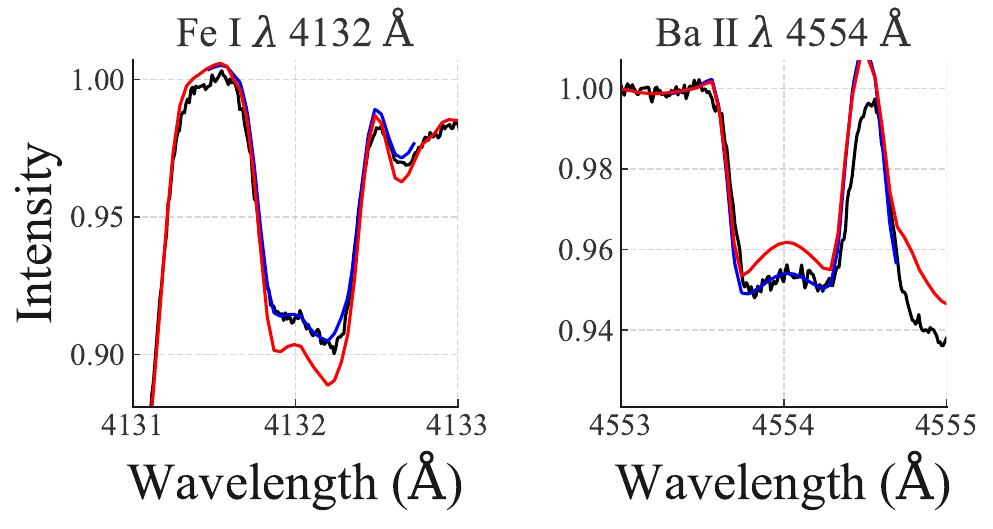}
    \caption{Two of the metallic line profiles of $\beta$ PsA used to determine its stellar properties. The black is the observed spectrum with the best-fit unscaled model in red and the scaled model in blue. Scaling the line profiles results in better fits.}
    \label{fig:scaledvunscaled}
\end{figure}

The stable atmospheres of spectral type A stars make them prone to peculiar abundances because of gravitational settling (e.g., see Section 1 in \citealt{xiang_chemically_2020}). We account for possible abundance variations by scaling the depth of each line separately. Omitting this step leads to line shapes that are overly distorted, biasing the stellar properties. As shown in Figure \ref{fig:scaledvunscaled}, the scaled best-fit model more accurately models the shapes of the lines, which is especially evident in the \ion{Fe}{1} and \ion{Ba}{2} lines. The scaling of the line profiles is conducted by calculating the ratio of the observed line depths ($f_{obs}$) and model line depths ($f_{model}$) at the center of each of the 11 distorted spectral lines. This ratio ($f_{obs}$/$f_{model}$) is then applied as a multiplicative factor to the continuum-subtracted line profile so that the depth of the line center of the models matches the observed spectrum. The continuum is reset to 1.0 before the line fitting analysis is conducted. 

We compute $\chi^2$ values for the 11 line profiles fit for each model. The average of these values for an individual model is referred to as $\chi^2_{shape}$. Poisson statistics are assumed for the uncertainties in the flux, and the average value weights each metallic line equally; they all have similar S/N values.  

The line shapes and scaled line profiles do not offer strong constraints on the stellar temperature. High temperatures at low inclinations provide similar quality fits as cool stars at high inclinations. To break this degeneracy, we measure the line ratios of six pairs of iron absorption lines, listed in Table \ref{tab:fe_ratios}. Line strengths of the observed and model spectra are determined by measuring the equivalent width (EW) of each line over a 1~\AA\ window using the \texttt{specutils} Python package. The uncertainty in the observed EW measurements are calculated using a standard formula, discussed in Appendix \ref{app:ew_uncertainty},

\begin{equation}\label{eq:sigma_EW}
    \sigma_{EW} = \frac{\sqrt{f_{obs}~\rm{FWHM}(\Delta\lambda)}}{\rm{SNR}}.
\end{equation}

\noindent Here, $f_{obs}$ is a flux term that scales the S/N of the continuum to the line depth. 
The difference between the model EW ratio and the observed EW ratio is determined for each line pair and for each synthetic spectrum in the grid, as:

\begin{equation}\label{eq:d_Fe}
    \Delta R_{ratio} = \frac{EW_{\mathrm{FeII_{Observed}}}}{EW_{\mathrm{FeI_{Observed}}}} - \frac{EW_{\mathrm{FeII_{Model}}}}{EW_{\mathrm{FeI_{Model}}}}.
\end{equation}

\noindent The uncertainty in each ratio is then propagated such that,

\begin{equation}\label{eq:sigma_Fe}
    \sigma_{ratio} = \sqrt{\Big(\frac{\sigma_{\rm{FeII}}}{EW_{\rm{FeI}}}\Big)^2+\Big(\sigma_{\rm{FeI}}\frac{EW_{\rm{FeII}}}{EW_{\rm{FeI}}^2}\Big)^2}.
\end{equation}

\noindent The uncertainty in the EWs for the model spectra are assumed to be zero. Table \ref{tab:fe_ratios} lists the observed EW ratios and the EW ratios for the best-fit model.

We compute a $\chi^2$ metric to measure the agreement of line ratios ($\chi^2_{ratio}$), computed using the following formula:

\begin{equation}\label{eq:chi_ratio}
    \chi^2_{ratio} = \sum_i \Big(\frac{\Delta R_{ratio,i}}{\sigma_{ratio,i}}\Big)^2.
\end{equation}

\noindent The best-fit synthetic spectrum of $\beta$ PsA is determined by minimizing a ``combined $\chi^2$" ($\chi^2_{comb}$), calculated as a quadrature sum of the $\chi^2$ from the line shape ($\chi^2_{shape}$) and the $\chi^2$ from the line ratios ($\chi^2_{ratio}$):

\begin{equation}\label{eq:combined_chi}
    \chi^2_{comb} = \frac{\chi^2_{shape}}{\chi^2_{shape,\rm{min}}} +  \frac{\chi^2_{ratio}}{\chi^2_{ratio,\rm{min}}}.
\end{equation}

\noindent We normalize the calculated $\chi^2$ values by the minimum $\chi^2$ value for both terms so that they are weighted equally. 

\begin{deluxetable}{cccc}
\tablecaption{6 pairs of Fe line EWs and \ion{Fe}{2}/\ion{Fe}{1} ratios \label{tab:fe_ratios}}
\tablehead{
\colhead{\ion{Fe}{2} line} & \colhead{\ion{Fe}{1} line} & \colhead{$EW_{\rm{Fe~II}}$/$EW_{\rm{Fe~I}}$} & \colhead{$EW_{\rm{Fe~II}}$/$EW_{\rm{Fe~I}}$}
\\
\colhead{\AA} & \colhead{\AA} & \colhead{Observed} & \colhead{Best-fit model}
}
\startdata
4451.5 & 4447.7 & $1.377\pm0.007$ & 1.693\\
4472.9 & 4476.1 & $0.943\pm0.003$ & 0.776\\
4491.4 & 4494.6 & $3.138\pm0.010$ & 3.323\\
4522.6 & 4528.6 & $2.483\pm0.004$ & 2.786\\
4731.5 & 4736.7 & $2.727\pm0.012$ & 2.595\\
4923.9 & 4920.5 & $2.427\pm0.003$ & 2.127\\
\enddata
\end{deluxetable}

\begin{deluxetable}{lrr}
\tablecaption{Best-Fit Model Parameters for $\beta$ PsA \label{tab:params}}
\tablehead{
\colhead{Input Parameters} & \colhead{This Work} & \colhead{Other Works}
}
\startdata
Inclination, $i$ (degrees) & $4.75^{+0.75}_{-0.50}$ & \\
Stellar Mass (M$_{\odot}$) & $2.20\pm0.03$ & $2.34\pm0.01^{a}$ \\
Polar Temperature (K) & $10300^{+200}_{-250}$ & \\
Equatorial Radius (R$_{\odot}$) & $2.20\pm0.21$ & \\
$\omega~(\Omega/\Omega_{k})$ & $0.72^{+0.09}_{-0.10}$ & \\
Metallicity ([M/H]) & $+0.0^{*}$ & $+0.01^{c}$ \\
\hline
~~~~Derived Parameters & & \\ 
\hline
$R_{\mathrm{eq}}/R_{\mathrm{pole}}$ & $1.26\pm0.07$ & \\
$T_{\mathrm{eff}}$ (K) & $9389^{+91}_{-133}$ & $9462\pm109^{a}$ \\ 
$T_{\mathrm{eq}}$ (K) & $8275^{+317}_{-400}$ & \\
$T_{\mathrm{eq}}/T_{\mathrm{pole}}$ & $0.80\pm0.04$ & \\
$L$ (L$_\odot$) & $26.2^{+1.9}_{-2.4}$ & $38.57\pm0.95^{a}$ \\
$\log g_{\mathrm{eq}}$ & $3.78^{+0.10}_{-0.13}$ & \\
$\log g_{\mathrm{pole}}$ & $4.29\pm0.05$ & \\
$\log g_{\mathrm{eff}}$ & $4.21\pm0.03$ & $4.20\pm0.01^{c}$ \\
$v_{\mathrm{eq}}$ (km s$^{-1}$) & $314^{+37}_{-43}$ & \\
$v_{\mathrm{eq}}~\sin~i$ (km s$^{-1}$) & $26.0^{*}$ & $26.0^{b}$ \\
$\Omega/\Omega_{\mathrm{crit}}$ & $0.93\pm0.17$ & \\
\enddata
\tablecomments{The values presented here are derived from the analysis of the HARPS spectrum.
$a$ — \cite{zorec_rotational_2012}, 
$b$ — \cite{glebocki_systematic_2005}, 
$c$ — \cite{soubiran_pastel_2016}, 
$^*$ — adopted values not fit
}
\end{deluxetable}

\begin{figure}
    \centering
    \includegraphics[width=\linewidth]{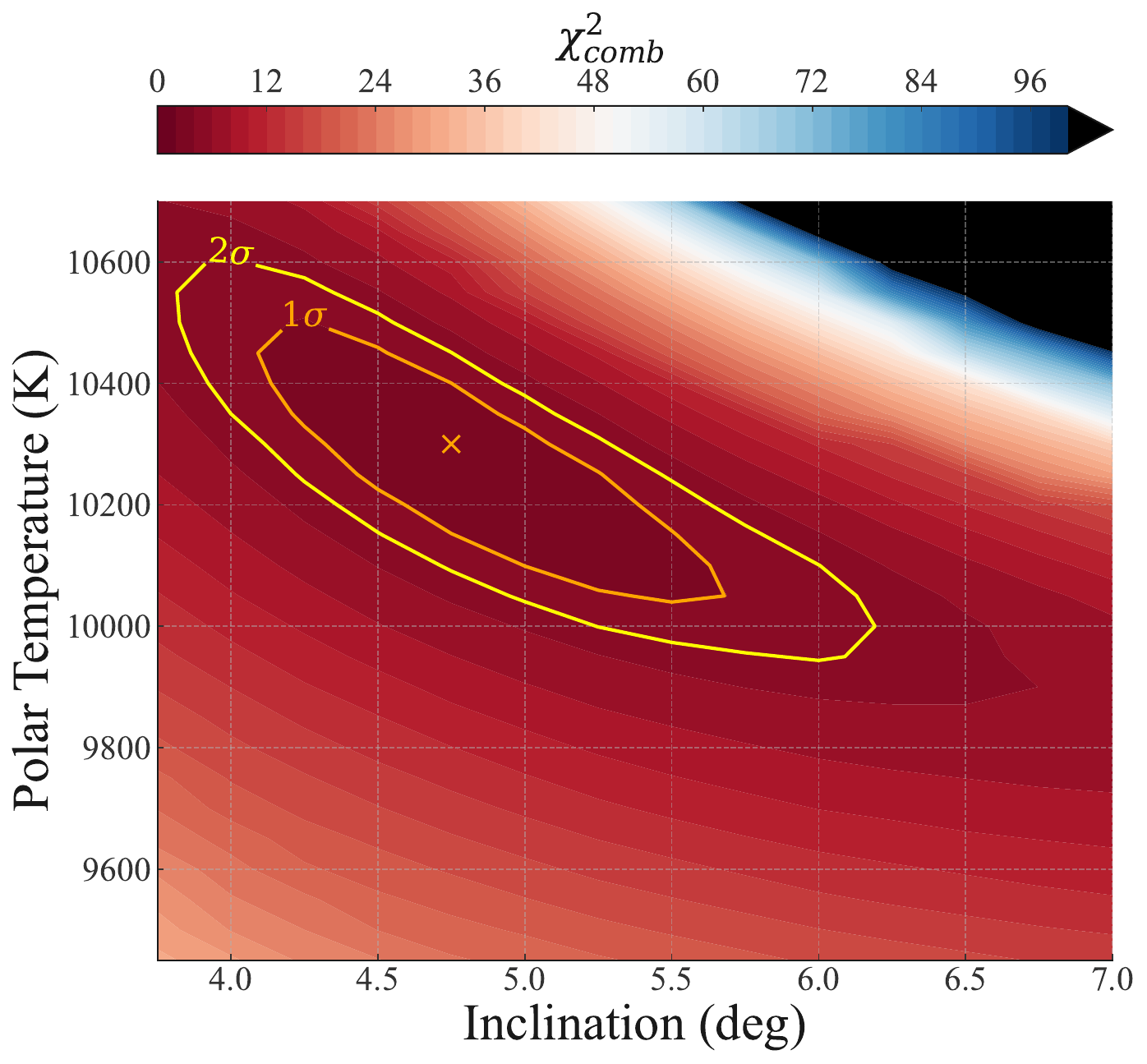}
    \caption{Contour plot of $\chi_{comb}^2$ values versus polar temperature and inclination. The ``x" marks the best-fit polar temperature and inclination. The orange curve traces the $1\sigma$ contour and the yellow curve traces the $2\sigma$ contour.}
    \label{fig:final_contour_2.20}
\end{figure}

\begin{figure*}[t]
    \centering
    \includegraphics[width=\linewidth]{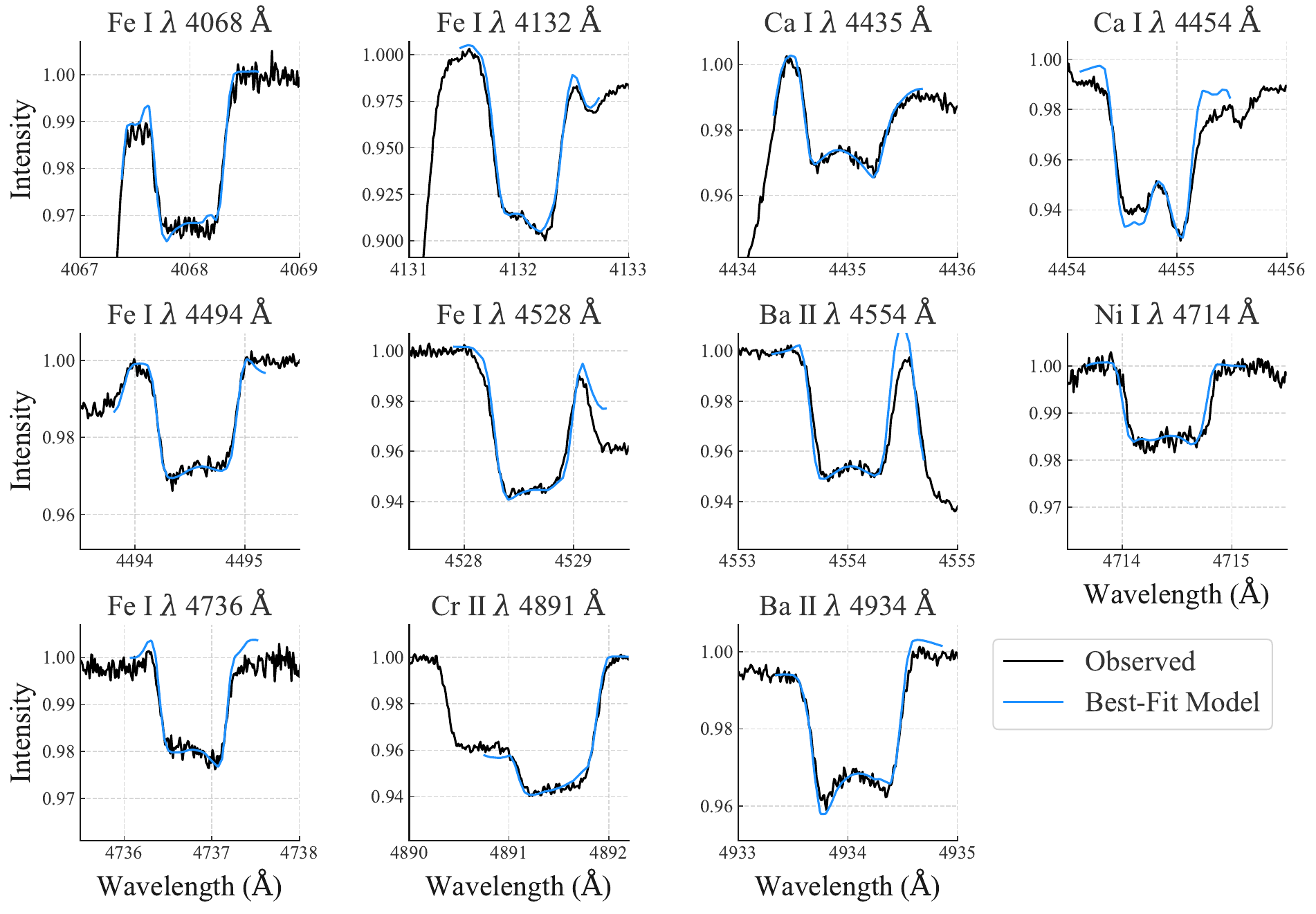}
    \caption{The 11 metallic line profiles in the HARPS spectrum of $\beta$ PsA (black) used to determine the best-fit model (blue).}
    \label{fig:best_fit}
\end{figure*}

The uncertainties in the polar temperature and inclination values are determined by varying each parameter until $\chi^2_{comb}$ increases by 1.0. The uncertainties for the other stellar parameters (i.e., $T_{\rm{eff}},~L,~log~g_{\rm{eff}},~v_{\rm{eq}}$, etc.) presented here are derived by finding the extreme values of each parameter from the models that are in the $1\sigma$ uncertainty in inclination and polar temperature. Figure \ref{fig:final_contour_2.20} shows the best-fit polar temperature and inclination, along with the $1\sigma$ range of these values. The best-fit model spectrum and observed HARPS spectrum are plotted in Figure \ref{fig:best_fit}.

\subsection{Verifying the Spectral Model at UV Wavelengths}

\begin{figure}
    \centering
    \includegraphics[width=\linewidth]{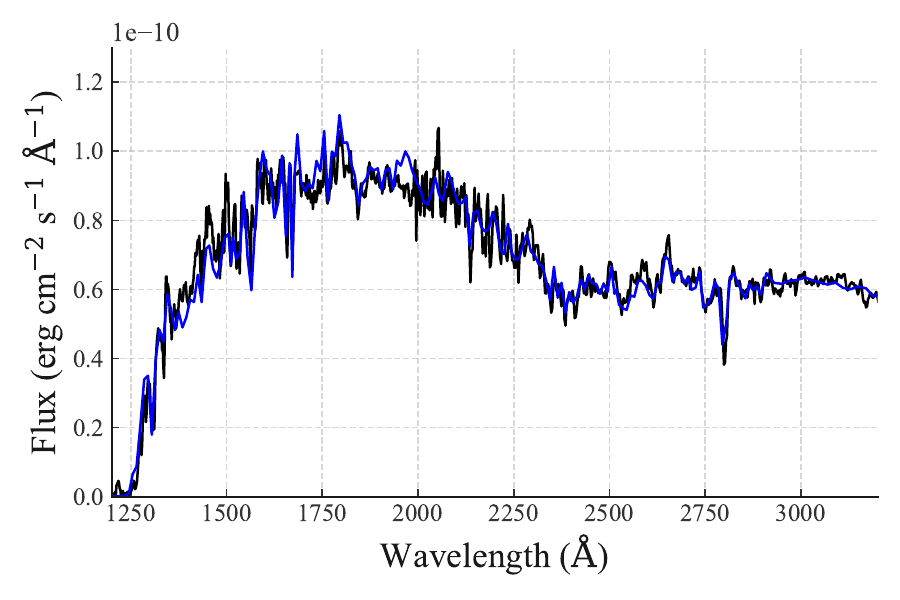}
    \caption{A merged IUE ultraviolet spectrum of $\beta$ PsA (black) is constructed from a scaled SWP spectrum and a cleaned LWP spectrum (see text). The best-fit synthetic spectrum (blue) normalized to the IUE continuum and computed using the low-resolution version of \texttt{fastrot-spec} is shown for comparison. The excellent agreement between the observations and the model confirms the quality of the fit and further validates the determined atmospheric properties.} \label{fig:UV_cont}
\end{figure}

We investigate the agreement of the synthetic spectrum of $\beta$ PsA with UV observations; a comparison at shorter wavelengths provides a stricter test of the multi-temperature, best-fit stellar model. Two low-resolution large-aperture spectra of $\beta$ PsA, obtained with the International Ultraviolet Explorer (\textit{IUE}; \citealt{boggess_history_1987}), are available: SWP42704 (1150--1980 \AA) and LWP21480 (1850--3200 \AA). In both the MAST\footnote{\url{https://archive.stsci.edu/iue/search.php}} and INES\footnote{\url{https://sdc.cab.inta-csic.es/ines/}} archival spectra, a mismatch in the flux calibration between the short- and long-wavelength spectra is apparent. For the analysis presented here, SWP42704 is scaled to the continuum level of LWP21480 by dividing its flux by a factor of 1.23.  In addition, the long-wavelength spectrum is cleaned of very noisy data in the 1850--1965 \AA\ interval. Figure \ref{fig:UV_cont} shows the merged spectrum obtained from the scaled SWP observations and the cleaned LWP  observations. Also shown is the synthetic spectrum computed with the low-resolution version of \texttt{fastrot-spec} using the parameters listed in Table~\ref{tab:params}; the synthetic spectrum has been normalized to the long-wavelength ultraviolet continuum. The agreement is excellent, further confirming the quality of the fit and the determined atmospheric parameters. 

\subsection{$\beta$ PsA is a Rapidly Rotating Star Seen Nearly Pole-On} \label{sec:results}

The determined stellar properties based on the spectral analysis of $\beta$ PsA are presented in Table \ref{tab:params}. Uncertainties are calculated for all determined parameters, but here we summarize the best-fit values. Despite its low \textit{v}~sin~\textit{i} value, $\beta$ PsA rotates with an equatorial velocity of $314$ km~s$^{-1}$, corresponding to $93\%$ of its break-up velocity. It is viewed nearly pole-on with an inclination of $4.75~^\mathrm{o}$. With an equatorial radius of $2.20$ R$_{\odot}$ and stellar mass of $2.20$ M$_\odot$, the rapid rotation causes an oblateness ($R_{\mathrm{eq}}/R_{\mathrm{pole}}$) of $1.26$. The gravity darkening associated with this results in the equatorial temperature ($8275$ K) being 2025 K cooler than the polar temperature ($10300$ K). This corresponds to a temperature ratio ($T_{\mathrm{eq}}/T_{\mathrm{pole}}$) of $0.80$.

\section{An Independent Analysis Using CHIRON Spectra}\label{sec:chiron}

\begin{figure}
    \centering
    \includegraphics[width=0.95\linewidth]{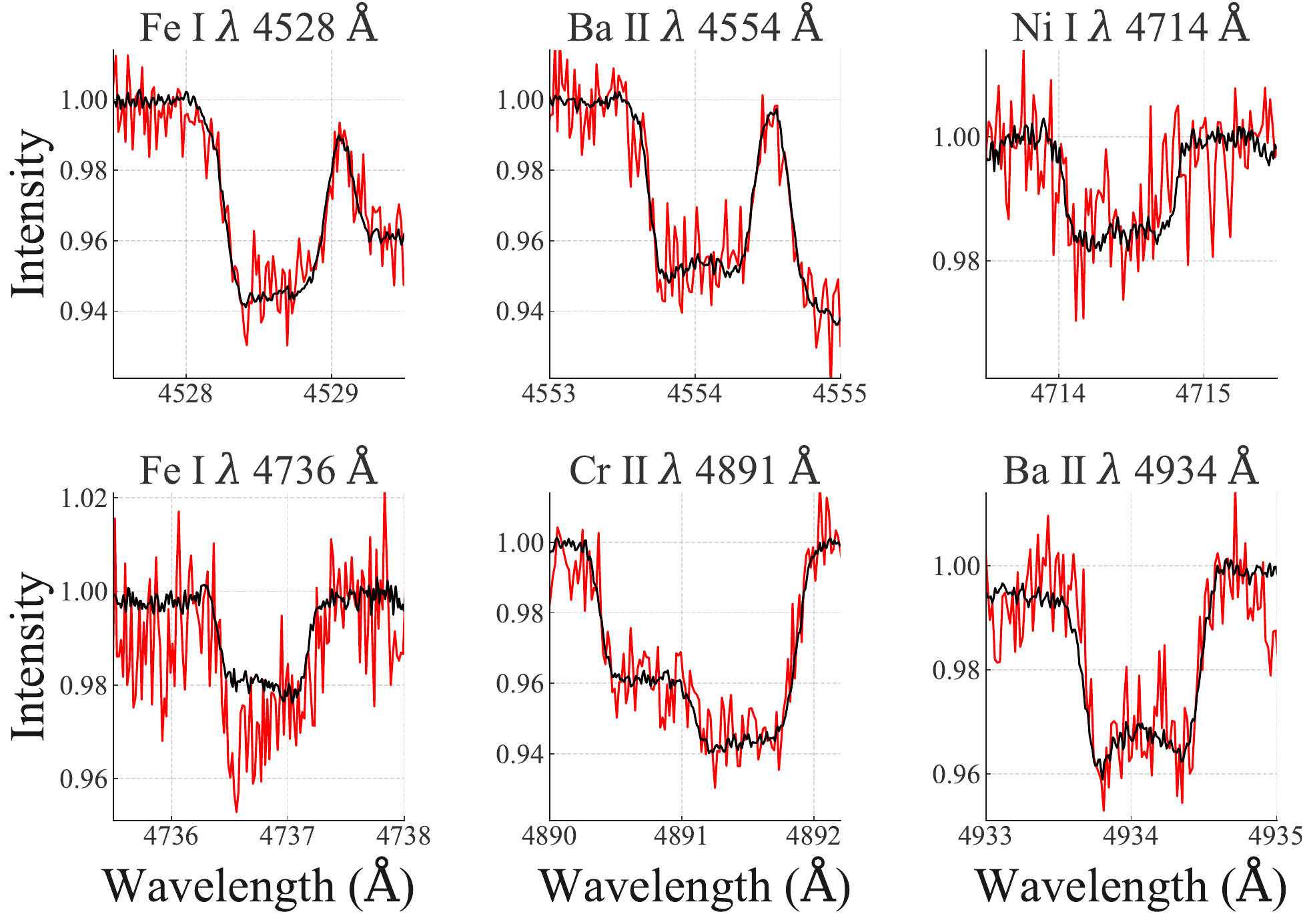}
    \caption{The 6 metallic line profiles of $\beta$ PsA used to determine its stellar properties in the CHIRON spectrum (red)  and the HARPS stacked spectrum (black).}
    \label{fig:chironvharps}
\end{figure}

\begin{figure}
    \centering
    \includegraphics[width=0.95\linewidth]{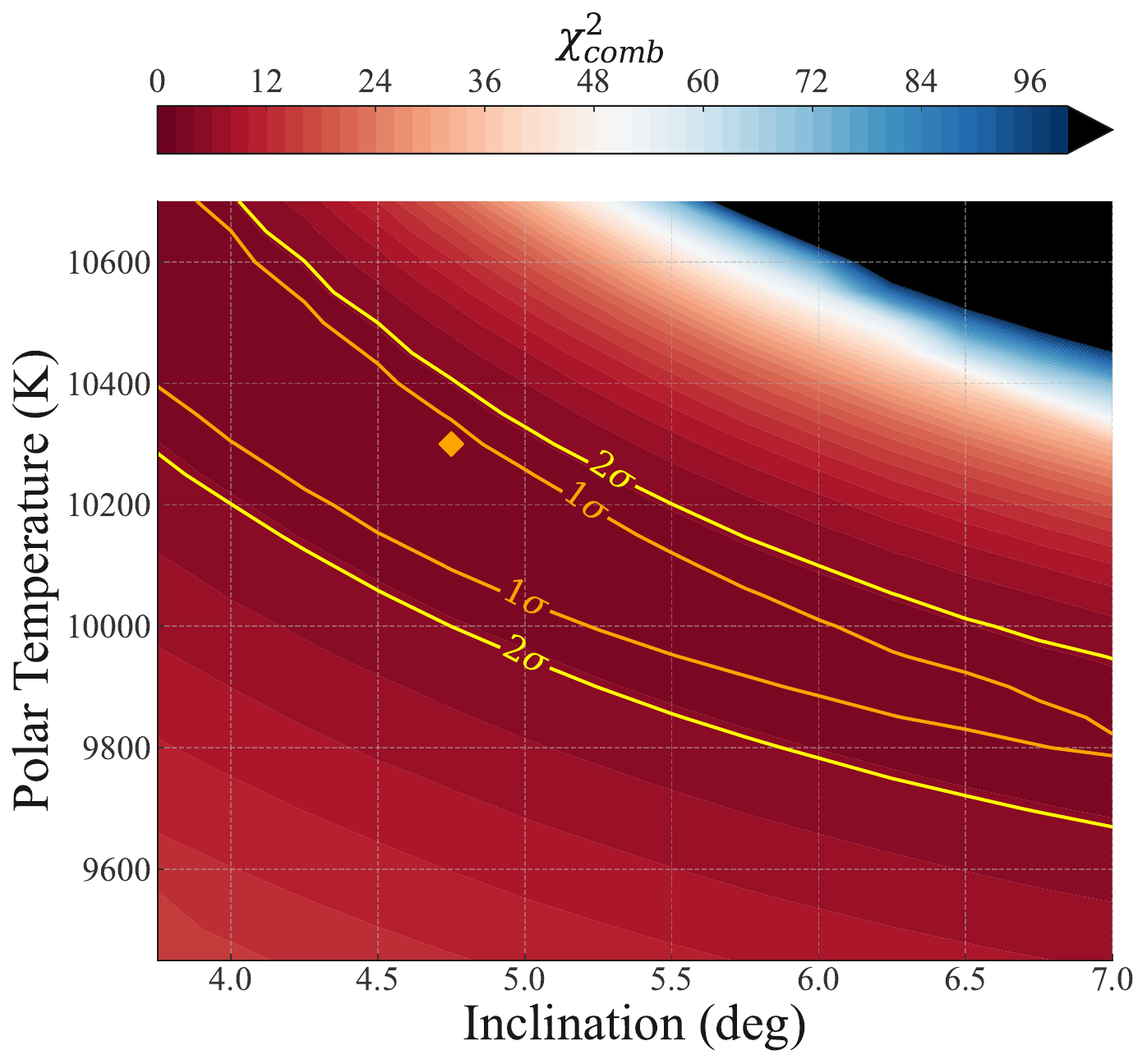}
    \caption{Contour plot of $\chi_{comb}^2$ values versus polar temperature and inclination for the CHIRON analysis. The $1\sigma$ (orange) and $2\sigma$ (yellow) contours are labeled. The best-fit values determined from the HARPS spectra (orange diamond) are consistent with the CHIRON results.}
    \label{fig:chiron_contour}
\end{figure}

An identical spectral modeling analysis is conducted using the CHIRON spectra of $\beta$ PsA presented in Section \ref{subsec:CHIRON}. This provides a check on systematic uncertainties inherent to the spectrograph used. Figure \ref{fig:chironvharps} shows both the HARPS (S/N $\sim980$) and CHIRON (S/N $\sim150$) spectra for comparison. The same radius (2.20 R$_{\odot}$) and final stellar mass (2.20 M$_{\odot}$) are used to compute the \texttt{fastrot-spec} synthetic spectra. 
Since CHIRON spectra do not extend blueward of 4500 \AA, only six of the 11 metallic line profiles and three of the six Fe line ratios are available. Because of the lower S/N, the saddle-shaped profiles of the lines cannot be modeled as precisely as in the high S/N spectrum from HARPS, so the $\chi^2_{comb}$ is more sensitive to the fit to line ratios ($\chi^2_{ratio}$) than the line ($\chi^2_{shape}$) and the best-fit model is skewed slightly toward a more pole-on, hotter model ($i = 3.75^{~\rm{o}}, T_{pole} = 10600$) which is within the $3\sigma$ uncertainty derived from the HARPS spectrum. The best-fit model derived from minimizing the $\chi^2_{shape}$ only using the CHIRON spectrum has an $i = 4.75^{~\rm{o}}$ and $T_{pole} = 10000$ K, which is consistent with the HARPS results. Figure \ref{fig:chiron_contour} shows the $\chi^2_{comb}$ contours for the CHIRON $\beta$ PsA spectrum. The $1\sigma$ contour from the CHIRON spectrum is substantially larger than from the HARPS spectrum. Uncertainties for the stellar parameters cannot be reliably ascertained for the CHIRON analysis due to the $1\sigma$ uncertainty extending beyond the parameter space that we are sampling here. Nevertheless, the relative agreement of the best-fit models with two independent spectra show that there are no large systematic errors in this analysis.

\section{Apparent Luminosity and Effective Temperature of Rapid Rotators}\label{sec:LandT}

Standard techniques used for computing luminosities rely on knowing the distance and some measure of the bolometric flux. Inherent to this is the assumption that the star has a homogeneous temperature and brightness across its surface. This is not true for rapidly rotating stars like $\beta$ PsA, and other rapidly rotating stars like $\beta$ Cas and $\alpha$ Leo (e.g., see Figures 6 and 7 in \citealt{che_colder_2011}); oblate stars do not emit the same flux in all directions. In this case, a star's ``apparent luminosity" can be significantly biased relative to its luminosity. For simplicity, we refer to the total energy output of a star per unit time as luminosity. For a nearly pole-on rapidly rotating star like $\beta$ PsA, its apparent luminosity will be much higher than its luminosity since we are primarily observing the higher temperature polar region.  The best-fit model confirms this. The best-fit model finds a luminosity of $26.2^{+1.9}_{-2.4}~L_{\odot}$ that is $48\%$ less than the luminosity measured by \cite{zorec_rotational_2012} ($38.6\pm1.0$ $L_\odot$).

A similar bias occurs when considering the effective temperature of a star. For a star with a nearly edge-on orientation ($i = 90^{\mathrm{o}}$), the cooler equatorial regions will dominate, making the star appear cooler than it actually is, while the opposite occurs for a nearly pole-on star. We calculate a mean-radius effective temperature (Section \ref{subsec:init_params}) from the models as a proxy for the effective temperature ($T_{\mathrm{eff}}$) calculated in stellar evolutionary models. This value is listed in Table \ref{tab:params} along with the other derived stellar parameters.

\section{The Luminosity, Mass, and Age of $\beta$ PsA}\label{sec:evolution}

We use $\beta$ PsA's luminosity and mean-radius effective temperature to estimate its mass and age via comparisons with PARSEC v2.0 mass tracks that account for rotation \citep{costa_multiple_2019, costa_mixing_2019, nguyen_parsec_2022,nguyen_parsec_2025} with a $\Omega/\Omega_{\mathrm{crit}}$ value of 0.95, where critical angular velocity is defined as, 

\begin{equation}
    \Omega_{\rm{crit}} = \sqrt{\frac{8}{27}\frac{GM}{R_{\rm{pole}}^3}}.
\end{equation} 

\begin{figure}
    \centering
    \includegraphics[width=\linewidth]{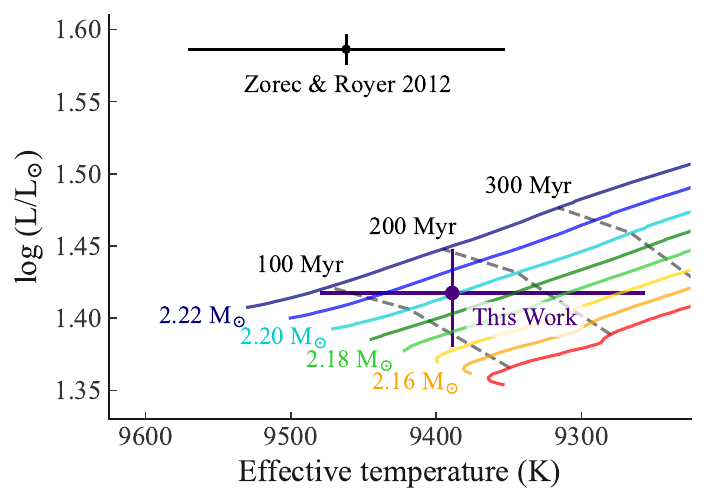}
    \caption{The mean-radius effective temperature and luminosity of $\beta$ PsA (indigo) are plotted on an HR diagram, along with PARSEC stellar evolutionary models with $\Omega/\Omega_{\mathrm{crit}} = 0.95$. The apparent effective temperature and luminosity from \cite{zorec_rotational_2012} (black) are also shown; the higher apparent luminosity arises from our nearly pole-on view of this rapidly rotating, gravity-darkened star.} 
    \label{fig:HR_diagram}
\end{figure}

\noindent This differs from the Keplerian velocity that is used to define the oblate geometry in the \cite{espinosa_lara_gravity_2011} parameterization, resulting in $\Omega/\Omega_{\rm{crit}} \ne \Omega/\Omega_k$. Our original best-fit model has a $\Omega/\Omega_{k}$ value of 0.72, which corresponds to a $\Omega/\Omega_{\mathrm{crit}} = 0.93 \pm0.17$. We use the \texttt{NumPy} \texttt{interpolate} function to linearly interpolate the mass tracks from 0.2~$M_\odot$ intervals to 0.01~$M_\odot$ intervals. These comparisons yield a mass of 2.20$\pm0.03$~M$_{\odot}$, with the uncertainty set by the uncertainties in the luminosity and the mean-radius effective temperature (see Figure \ref{fig:HR_diagram}). These models yield an age of $141^{+113}_{-49}$~Myr for $\beta$ PsA. This age is consistent with the age of $180\pm90$~Myr, reported by \cite{vican_age_2012}. Adopting the \cite{zorec_rotational_2012} values for $T_{\rm eff}$ and apparent $L$ and applying the same PARSEC models yields a mass that is 5\% larger (2.31~$M_\odot$) and an age that is 163\% larger (371~Myr) than those obtained here. This discrepancy highlights the importance of accounting for rapid rotation, even for stars that appear to be slow rotators (i.e., low $\sin i$).

The CHIRON spectrum of the companion star, CD-32~17127, shows strong lithium absorption of the blended $\lambda6708~\rm{\AA}$ feature, a well-known signature of youth. The EW of the total Li doublet absorption is determined by modeling each component with Voigt profiles; the model also accounts for the \ion{Fe}{1} line at  $\lambda$6707.44 \AA. This yields a \ion{Li}{1}~6708~\AA\ EW of $102.4\pm1.1$ m\AA.

The EW of the \ion{Li}{1} doublet can be used to estimate a Li-depletion age, once the effective temperature of the star is known. We determine the effective temperature of CD-32 17127 by using \texttt{SpecMatch-Emp} \citep{yee_precision_2017} as implemented on CHIRON data in \cite{hubbard-james_solar_2026}. This yields an effective temperature of $5964\pm110$ K. This effective temperature and Li EW are then compared to the Estimating AGes from Lithium Equivalent widthS (EAGLES; \citealt{jeffries_gaia_2023}) Li depletion models. This comparison yields an age of $416.9^{+434.3}_{-258.4}$~Myr. This age is consistent to within $1\sigma$ of the age of $\beta$ PsA (141~Myr) when the uncertainties are combined.

\section{A Validation Test Using the Star Vega}\label{sec:vega}

To validate our newly developed methodology for determining the inclination of a nearly pole-on rapidly rotating star, we apply this technique to the star Vega. 
We use the same methods detailed in Section \ref{sec:modeling} for determining the stellar properties of Vega. The spectrum of Vega used in this analysis is the HIDES spectrum provided by \cite{takeda_high-resolution_2007}, introduced in Section \ref{subsec:HIDES}. The stellar mass ($M$) and equatorial radius ($R_{\mathrm{eq}}$) are taken from \cite{monnier_imaging_2007} with $v$~sin~$i$ taken from \cite{takeda_determination_2021}. A grid of models with parameters listed in Table \ref{tab:model_grid} is computed, and the same $\chi^2_{comb}$ analysis is conducted to determine the best-fit. We note that \texttt{fastrot-spec} code was developed to model the spectrum of Vega using interferometrically measured properties; the spectral synthesis was not tuned to optimize a fit, as is done here.

\begin{figure}
    \centering
    \includegraphics[width=\linewidth]{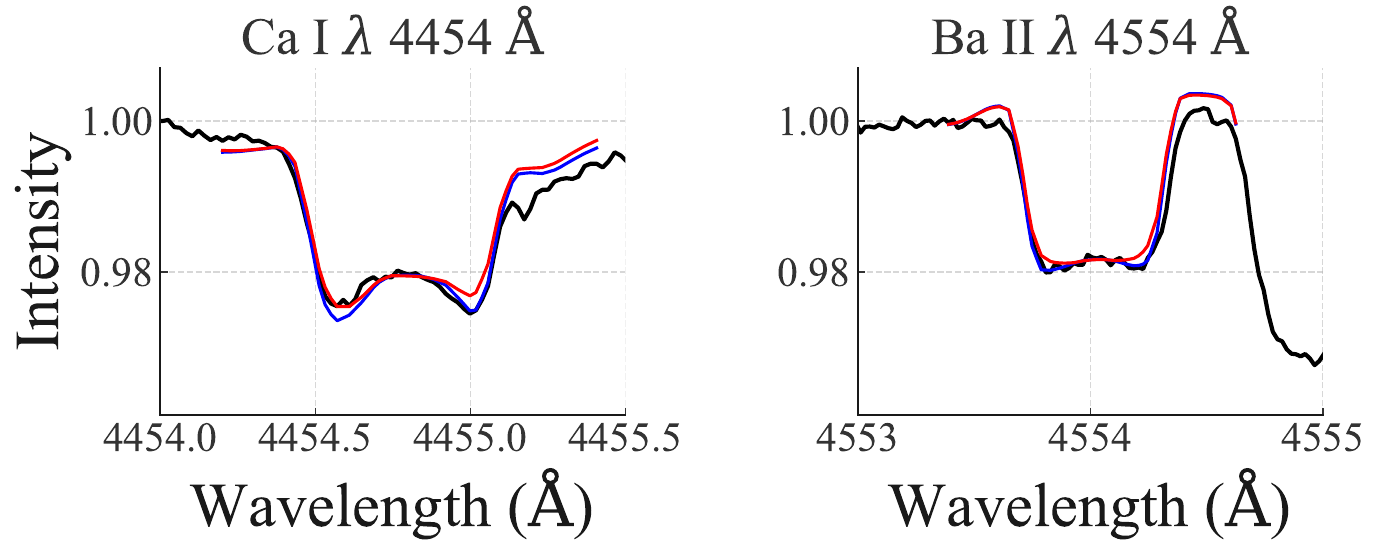}
    \includegraphics[width=\linewidth]{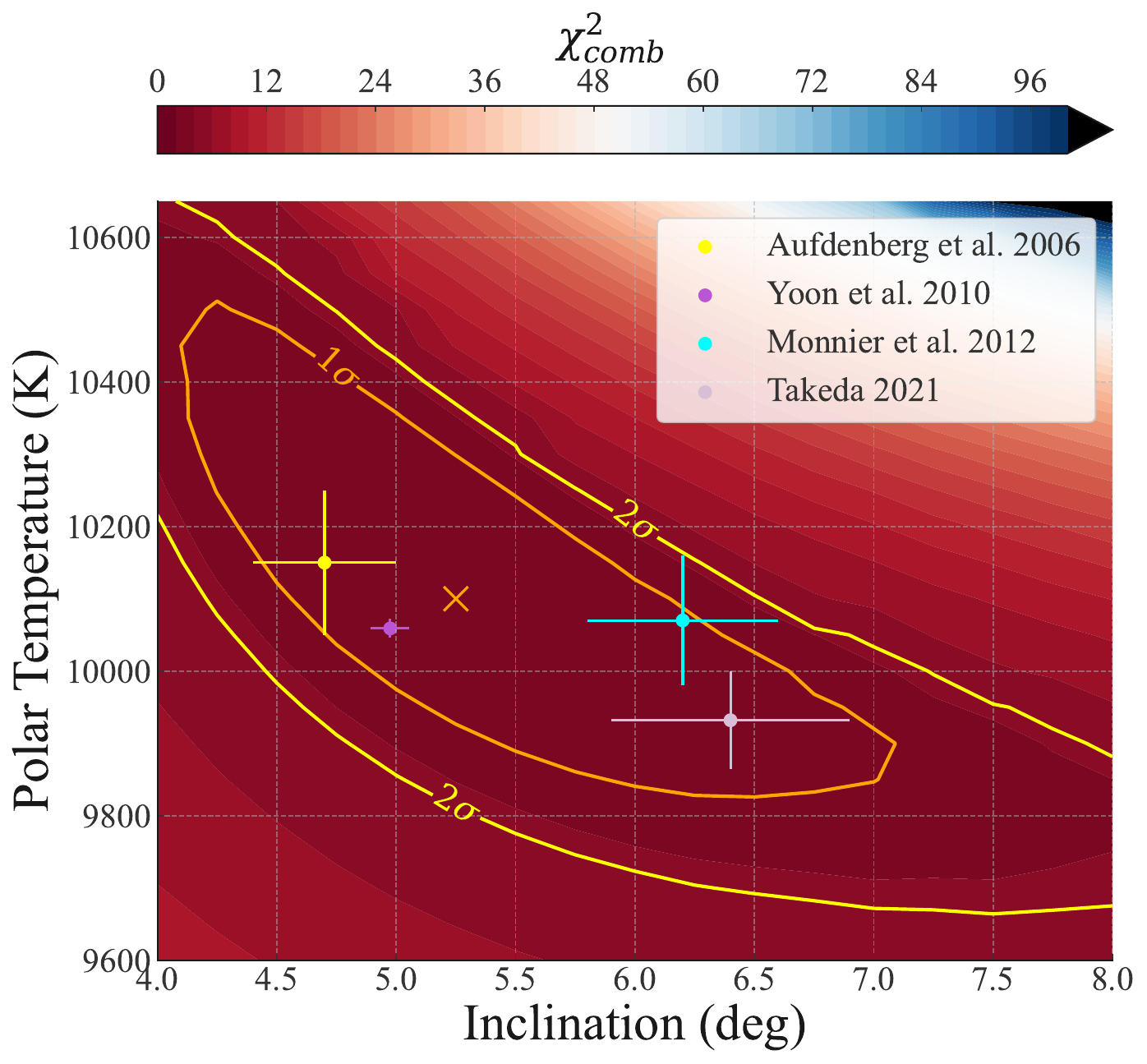}
    \caption{(\textbf{Top}) 2 of the 11 spectral lines modeled in the \cite{takeda_high-resolution_2007} Vega spectrum (black) with the best-fit model from this work (blue) and the original model from \citet[][red]{montesinos_surface_2024}. (\textbf{Bottom}) $\chi_{comb}^2$ contour for the analysis of the \cite{takeda_high-resolution_2007} spectrum of Vega. Our best-fit contours are in agreement with literature values for the polar temperature and inclination of Vega from \cite{aufdenberg_first_2006,yoon_new_2010, monnier_resolving_2012} \& \cite{takeda_determination_2021} in yellow, purple, cyan, and pink, respectively.}
    \label{fig:vega_plots}
\end{figure}

The best-fit model for Vega returns an inclination of $5.25^{+1.75}_{-1.00}$~$^{\circ}$ and a polar temperature of $10100^{+400}_{-250}$~K, which are consistent with the literature values of $i~=~6.2~^{\circ}$ and $T_{\mathrm{pole}} = 10000$~K \citep{monnier_resolving_2012,montesinos_surface_2024}. Table \ref{tab:vega_params} lists all the model parameters derived from this analysis, with values from other works shown for comparison. Two modeled spectral lines are shown in Figure \ref{fig:vega_plots} (top panel). The $\chi^2_{comb}$ contour in the polar temperature versus inclination plane is shown in Figure \ref{fig:vega_plots} (bottom panel) with literature values from \cite{aufdenberg_first_2006}, \cite{yoon_new_2010}, \cite{monnier_resolving_2012}, and \cite{takeda_determination_2021} shown for comparison. Our technique yields a result consistent with previous studies, and the $1\sigma$ contour conveys the correlation between inclination and polar temperature. When comparing the best-fit model to PARSEC evolutionary models with an appropriate value for $\Omega/\Omega_{crit}$ of 0.9, the mass derived is slightly higher ($\simeq$ 2.35 M$_{\odot}$), but within 10$\%$ of the mass determined by \cite{monnier_resolving_2012} and adopted as an initial parameter here.

\begin{deluxetable}{lrr}[h!]
\tablecaption{Best-Fit Model Parameters for Vega \label{tab:vega_params}}
\tablehead{\colhead{Input Parameters} & \colhead{This Work} & \colhead{Other Works}} 
\startdata
Inclination, $i$ (degrees) & $5.25^{+1.75}_{-1.00}$ & $6.2\pm0.4^{a}$\\
Stellar Mass ($\mathrm{M}_{\odot}$) & 2.15$^*$ & $2.15^{+0.10}_{-0.15}$$^{a}$ \\
Polar Temperature (K) & $10100^{+400}_{-250}$ & $10070\pm90^{a}$ \\
Equatorial Radius ($\mathrm{R}_{\odot}$) & $2.726^{*}$ & $2.726\pm0.006^{a}$\\
$\omega~(\Omega/\Omega_{k})$ & $0.60\pm0.15$ & $0.51^{b}$ \\
Metallicity ($[M/H]$) & $-0.5^{*}$ & $-0.5^{b}$\\
\hline
~~~~Derived Parameters & & \\ 
\hline
$\mathrm{R}_{\mathrm{eq}}/\mathrm{R}_{\mathrm{pole}}$ & $1.18^{+0.10}_{-0.08}$ & $1.13^{b}$\\
${T}_{\mathrm{eff}}$ (K) & $9375^{+186}_{-144}$ & $9360\pm90^{a}$ \\
$T_{\mathrm{eq}}$ (K) & $8624^{+392}_{-511}$ & $8910\pm130^{a}$\\
$T_{\mathrm{eq}}/T_{\mathrm{pole}}$ & $0.85\pm0.06$ & $0.89^{b}$\\
$\mathrm{L}$ (L$_{\odot}$) & $43.1^{+4.8}_{-5.1}$ & $47.2\pm2.0^{a}$ \\
log $g_{\mathrm{eq}}$ & $3.70^{+0.10}_{-0.15}$ & $3.769^{b}$\\
log $g_{\mathrm{pole}}$ & $4.04^{+0.07}_{-0.06}$ & $4.005^{b}$\\
log $g_{\mathrm{eff}}$ & $3.98\pm0.04$ & $3.958^{b}$ \\
$v_{\mathrm{eq}}$ (km s$^{-1}$) & $234^{+55}_{-58}$ & $195\pm15^{c}$ \\
$v_{\mathrm{eq}}~\sin~i$ (km s$^{-1}$) & $21.4^{*}$ & $21.6\pm0.3^{c}$ \\
$\Omega/\Omega_{\mathrm{crit}}$ & $0.86\pm0.31$ & $0.774\pm0.012^{a}$ \\
\enddata
\tablecomments{\textit{a} - \cite{monnier_resolving_2012} Model 3, \textit{b} - \cite{montesinos_surface_2024}, \textit{c} -\cite{takeda_determination_2021}, $^*$ - adopted values not fit.}
\end{deluxetable}

\section{Summary and Discussion}\label{sec:discussion} 

In this work, a methodology is developed for determining the inclination of the spectral type A1 main sequence star $\beta$ PsA using high spectral resolution (R~$=~115,000$) optical spectra from the HARPS spectrograph. In particular, 11 metallic lines are modeled to constrain the gravity darkening and inclination and six 
temperature-sensitive line ratios are used to constrain the absolute polar temperature. The analysis confirms that $\beta$ PsA is a rapidly rotating star ($\Omega/\Omega_{\rm crit}=0.93\pm0.17$) oriented nearly pole-on ($i = 4.75^{+0.75}_{-0.50}$$^{\circ}$). It has a polar temperature of $10300^{+200}_{-250}$ K that is 24\% hotter than its equatorial temperature ($8275^{+317}_{-400}$ K), causing its apparent luminosity to be 48\% larger than its luminosity of $26.2^{+1.9}_{-2.4}$ L$_\odot$. When its mean-radius effective temperature and luminosity are compared with stellar evolutionary models, they yield a mass of $2.20\pm0.03$ M$_{\odot}$ and an age of $141^{+113}_{-49}$~Myr. This youthful age is consistent with the detection of lithium in the spectrum of its common proper motion companion, the G5V star CD-32 17127. The inferred stellar properties of $\beta$ PsA are consistent with those measured using lower-resolution, lower-S/N spectra from the CHIRON spectrograph.

The effects of rotation on intermediate and high mass stars are important as they can alter mixing throughout the stellar interior and shift the course of evolution \citep{maeder_evolution_2000}. 
Constraining these processes is potentially possible by directly measuring the oblate shape and gravity darkening of rapidly rotating stars via long-baseline optical/infrared interferometry (e.g., \citealt{monnier_imaging_2007, zhao_imaging_2009, che_colder_2011, van_belle_interferometric_2012}). The work presented here demonstrates the potential to constrain the assumptions used in modeling rapidly rotating stars (e.g., gravity-darkening laws; \citealt{espinosa_lara_gravity_2011}) using spectroscopy alone.

The success of this analysis on lower S/N spectra obtained on a 1.5-meter telescope showcases the potential to extend this work to a larger scale. For example, a survey of low \textit{v}~sin~\textit{i} spectral type A stars could help distinguish between low $v_{\rm{eq}}$ and low sin~$i$, and thereby more clearly determine the rotation distribution of these stars (e.g., \citealt{zorec_rotational_2012}).  Additionally, under the assumption of low obliquity, the identification of such systems could help identify face-on debris disk systems and/or planetary systems. While face-on systems are the hardest to identify (e.g., \citealt{engler_characterization_2025}), they offer the best targets for understanding how planets dynamically interact and sculpt their resident disk (e.g., \citealt{farhat_case_2023,milli_alma_2026}).

\section{Acknowledgments}
The authors thank Doug Gies, Becky Flores, Mahir Patel, Akshat Chaturvedi, Ella Roselli, Jane Pratt, and Jack McGuire for helpful discussions, and Jenny Patience and Stanimir Metchev for insightful comments on the use of low-obliquity configurations as tracers of face-on debris disks and planetary systems. CK thanks Zach Way for guidance in implementing the \texttt{mdwarf-contin} code for the spectra used in this work. RW acknowledges support from the National Science Foundation Grant AST-2511438. BM acknowledges the funding by grant PID2021-127289-NB-I00 from MCIN/AEI/10.13039/501100011033/ and FEDER. We have used data from the CHIRON spectrograph on the SMARTS 1.5m telescope, which is operated as part of the SMARTS Observatory by RECONS\footnote{www.recons.org} members. Version 1.0.0 of the fastrot-spec code \citep{montesinos_fastrot-spec_2024} was used for the calculations in this paper. astrobmm/fastrot-spec is licensed under the GNU General Public License v3.0. The IUE data presented in this paper were obtained from the Mikulski Archive for Space Telescopes (MAST) at the Space Telescope Science Institute. The specific observations analyzed can be accessed via \dataset[https://doi.org/10.17909/6xkv-jg17]{https://doi.org/10.17909/6xkv-jg17}. STScI is operated by the Association of Universities for Research in Astronomy, Inc., under NASA contract NAS5–26555. Support to MAST for these data is provided by the NASA Office of Space Science via grant NAG5–7584 and by other grants and contracts.

\begin{contribution}
CK developed the methodology, prepared the data, led the analysis, and wrote the manuscript. RW advised CK and assisted in developing the methodology and conclusions for this work. JJ and BM contributed to the conceptual development of the project and provided critical review and feedback on the manuscript. SCG conducted the spectral analysis of CD-32 17127. SCG and WCJ scheduled the CHIRON observations. WCJ and TH provided access to the CHIRON instrument used in this study and supported data acquisition. AK conducted the observations of the CHIRON spectra, and TJ reduced the CHIRON spectra.
\end{contribution}

\facilities{ESO:3.6m (HARPS), CTIO:1.5m (CHIRON), OAO:1.88m (HIDES)}

\software{fastrot-spec \citep{montesinos_surface_2024, montesinos_fastrot-spec_2024}, Astropy \citep{astropy_collaboration_astropy_2013}, Specutils \citep{nicholas_earl_astropyspecutils_2024}, mdwarf-contin \citep{medan_importance_2025}, SpecMatch-Emp \citep{yee_precision_2017}, NumPy \citep{harris_array_2020}, PyAstronomy \citep{czesla_pya_2019}}

\appendix
\section{Equivalent-Width Uncertainties with Flux-Dependent Noise}
\label{app:ew_uncertainty}

The standard expression for the uncertainty in an equivalent-width (EW) measurement (e.g., \citealt{cayrel_data_1988, deliyannis_evidence_1993}) assumes that the signal-to-noise ratio (S/N) within an absorption line is identical to that of the surrounding continuum. This assumption is appropriate for weak lines (the line depths range from 0.94 to 0.99 for $\beta$ PsA) but can be relaxed for photon-noise–dominated spectra, where the noise scales with the square root of the detected flux. Here we derive a generalized expression that accounts explicitly for the reduced flux within an absorption line.

We assume that the spectrum is dominated by photon noise, such that the variance in each pixel is proportional to the detected flux. The spectrum is normalized to the continuum level, which is unity, and the reported signal-to-noise ratio, denoted $\mathrm{SNR}_c$, refers to the continuum. The normalized flux within the absorption line is denoted $f(\lambda)$, with $0 \le f(\lambda) \le 1$, where $f=1$ corresponds to the continuum and $f=0$ to complete absorption.

Under photon-noise domination, the flux uncertainty in pixel $i$ scales as the square root of the flux. In normalized units, the uncertainty in the flux is therefore

\begin{equation}
\sigma_{F,i}
=
\frac{\sqrt{f_i}}{\mathrm{SNR}_c}.
\end{equation}

For a pixel of width $\Delta\lambda$, the contribution of pixel $i$ to the equivalent width is

\begin{equation}
\mathrm{EW}_i = (1 - f_i)\,\Delta\lambda.
\end{equation}

The uncertainty in this contribution is given by propagation of the flux uncertainty,

\begin{equation}
\sigma_{\mathrm{EW},i}
=
\sigma_{F,i}\,\Delta\lambda
=
\frac{\sqrt{f_i}}{\mathrm{SNR}_c}\,\Delta\lambda.
\end{equation}

Assuming that the noise is uncorrelated between pixels, the total variance in the equivalent width is obtained by summing the pixel-level variances in quadrature:

\begin{equation}
\sigma_{\mathrm{EW}}^2
=
\sum_{i=1}^{N}
\sigma_{\mathrm{EW},i}^2
=
\frac{\Delta\lambda^2}{\mathrm{SNR}_c^2}
\sum_{i=1}^{N} f_i.
\end{equation}

Taking the square root yields

\begin{equation}
\sigma_{\mathrm{EW}}
=
\frac{\Delta\lambda}{\mathrm{SNR}_c}
\sqrt{\sum_{i=1}^{N} f_i}.
\end{equation}

In the limit of small pixel widths, the discrete sum may be approximated by an integral over the line profile:

\begin{equation}
\sum_i f_i
\;\approx\;
\frac{1}{\Delta\lambda}
\int_{\mathrm{line}} f(\lambda)\,d\lambda.
\end{equation}

Substituting this into the expression above gives

\begin{equation}
\sigma_{\mathrm{EW}}
=
\frac{1}{\mathrm{SNR}_c}
\sqrt{
\Delta\lambda
\int_{\mathrm{line}} f(\lambda)\,d\lambda
}.
\end{equation}

For a line of total width $W$ with approximately constant normalized flux $f$ across the line, the integral simplifies to

\begin{equation}
\int_{\mathrm{line}} f(\lambda)\,d\lambda = f\,W,
\end{equation}

yielding

\begin{equation}
\sigma_{\mathrm{EW}}
=
\frac{1}{\mathrm{SNR}_c}
\sqrt{f\,W\,\Delta\lambda}.
\end{equation}

This expression reduces to the familiar scaling in the limit $f \rightarrow 1$, appropriate for weak absorption lines, while explicitly accounting for reduced photon noise within deeper lines.

\newpage

\bibliography{reference}{}
\bibliographystyle{aasjournal}

\end{document}